# Climate Cycling on Early Mars Caused by the Carbonate-Silicate Cycle

Natasha E. Batalha[1,2,3*], Ravi Kumar Kopparapu[3,4,5,6], Jacob Haqq-Misra[3,4], and James F. Kasting[2,3,7]

[1]*Department of Astronomy & Astrophysics, Penn State University, University Park, PA 16802, USA*
[2] *Center for Exoplanets and Habitable Worlds, Pennsylvania State University, University Park, PA 16802, USA*
[3] *NASA Astrobiology Institute's Virtual Planetary Laboratory, P.O. Box 351580, Seattle, WA 98195, USA*
[4] *Blue Marble Space Institute of Science, 1001 4th Ave Suite 3201, Seattle, WA 98154*
[5] *NASA Goddard Space Flight Center, 8800 Greenbelt Road, Mail Stop 699.0 Building 34, Greenbelt, MD 20771, USA*
[6] *Department of Astronomy, University of Maryland, College Park, MD 20771, USA*
[7]*Department of Geosciences, Penn State University, University Park, PA 16802, USA*
[*]Corresponding Author. Email: neb149@psu.edu

**Abstract:** For decades, scientists have tried to explain the evidence for fluvial activity on early Mars, but a consensus has yet to emerge regarding the mechanism for producing it. One hypothesis suggests early Mars was warmed by a thick greenhouse atmosphere. Another suggests that early Mars was generally cold but was warmed occasionally by impacts or by episodes of enhanced volcanism. These latter hypotheses struggle to produce the amounts of rainfall needed to form the martian valleys, but are consistent with inferred low rates of weathering compared to Earth. Here, we provide a geophysical mechanism that could have induced cycles of glaciation and deglaciation on early Mars. Our model produces dramatic climate cycles with extended periods of glaciation punctuated by warm periods lasting up to 10 Myr—much longer than those generated in other episodic warming models. The cycles occur because stellar insolation was low, and because $CO_2$ outgassing is not able to keep pace with $CO_2$ consumption by silicate weathering followed by deposition of carbonates. While $CO_2$ by itself is not able to deglaciate early Mars in our model, we assume that the greenhouse effect is enhanced by substantial amounts of $H_2$ outgassed from Mars' reduced crust and mantle. Our hypothesis can be tested by future Mars exploration that better establishes the time scale for valley formation.

## 1. Introduction

Observational evidence for large-scale fluvial features on the martian surface dates back to NASA's Mariner 9 and Viking Missions (Masursky et al., 1977). Most of these features formed in the Late Noachian and Early Hesperian, ~3.8 Gyr ago, based on ages determined from crater counting (Fassett and Head, 2008). The geomorphological evidence includes the presence of valley networks (Cabrol and Grin, 2001), tributaries (Irwin et al., 2008), meandering channels (Hynek et al., 2010), open and closed-basin lakes (Goldspiel and Squyres, 1991), and the presence of phyllosilicates (Poulet et al., 2005). New findings from Mars Science Laboratory (MSL) indicate that Gale Crater was also once filled with liquid water for 10,000 to 10 million years (Grotzinger et al., 2015), implying that surface liquid water persisted for prolonged periods of time. Recent ground-based observations of deuterium-to-hydrogen (D/H) ratios in martian ground ice imply that at least 137 m of exchangeable water was originally available (Villanueva et al., 2015). The implied initial water inventory could be significantly larger if much of the H escaped hydrodynamically, carrying D with it (Batalha et al., 2015).

Despite the abundant evidence for sustained liquid water, climate modelers have had difficulty explaining how early Mars could have maintained a temperate environment. The Sun was ~25 percent less bright at the time when most of the valleys formed, making it difficult or impossible to warm the surface using only the greenhouse gases $CO_2$ and $H_2O$ (Kasting, 1991). Even 3-D climate models that include seasonal cycles and obliquity variations are unable to produce warm conditions (Forget et al., 2013). Adding $SO_2$ or $CH_4$ to the mix of greenhouse gases does not generate significant warming: the $SO_2$ either rains out (Batalha et al., 2015; Halevy and Head III, 2014) or photolyzes to form sulfate aerosols (Kerber et al., 2015; Tian et al., 2010), whereas $CH_4$ absorbs incoming solar near-infrared radiation in the stratosphere, creating anti-greenhouse cooling that offsets its greenhouse warming (Ramirez et al., 2014).

One supplementary greenhouse gas, $H_2$, can produce a warm climate, provided that it is supplied in sufficient quantities (Ramirez et al., 2014). $H_2$ absorbs effectively across the thermal-infrared spectrum as a consequence of collision-induced excitation of its rotational energy levels (Wordsworth and Pierrehumbert, 2013). 1-D climate modeling suggests that $CO_2$ partial pressures of >1.5 bar, combined with $H_2$ mixing ratios of 5 percent or more, could have kept early Mars' mean surface temperature above the freezing point of water (Ramirez et al., 2014). Outgassing of $H_2$ could result from a highly reduced mantle (Batalha et al., 2015; Ramirez et al., 2014) or from serpentinization of ultramafic crust (Batalha et al., 2015; Chassefière et al., 2014). Greenhouse warming by $H_2$ plays a critical role in the episodic warming mechanism described below.

In addition to $H_2$, the Ramirez et al. (2014) greenhouse warming mechanism also requires substantial amounts (1.5-3 bar) of $CO_2$. From the standpoint of initial planetary inventories, this amount of $CO_2$ is not unreasonable. Earth has the equivalent of ~60 bar of $CO_2$ tied up in carbonate rocks in its crust (Walker, 1985). If Mars was endowed with the same amount of carbon per unit mass, its initial inventory should have been on the order of 10 bars, accounting for its 9×smaller mass, 4×smaller surface area, and 3×smaller gravity. Questions remain as to how fast this carbon would have been outgassed and whether $CO_2$, or any atmospheric gas, could have been retained, given the intense solar EUV flux early in solar system history (Lammer et al., 2013). But if much of the carbon was sequestered in carbonates early on, then perhaps it could have been protected from loss during this time. Continued recycling of volatiles between the crust and the mantle would be required to revolatilize $CO_2$ later on and sustain significant volcanic outgassing rates, as discussed further below.

The scarcity of visible carbonate outcrops near Mars' surface has been used to argue that such a dense $CO_2$ atmosphere never existed (Hu et al., 2015). But carbonate has been observed in martian dust (Bandfield et al., 2003), in SNC meteorites (Bridges et al., 2001), and in the bottoms of fresh craters (Michalski and Niles, 2010), so it is definitely present within the crust. The lack of surface outcrops could be caused by the high acidity of rainwater under a dense $CO_2$ atmosphere, which may have dissolved carbonate minerals near the surface and reprecipitated them at depth (Kasting, 2010, Ch. 8).

Despite the evidence for fluvial activity, the difficulties in producing warm climates discussed above have spawned the idea that the early martian surface was generally frozen. Outflow channels on Mars have been interpreted as large flooding events, perhaps caused by magmatic heating of subsurface ice followed by breaching of groundwater through an icy surface (Baker, 1962; Cassanelli et al., 2015). Valley networks have been interpreted as snow migration, perhaps caused by obliquity variations and seasonal melting (Wordsworth et al., 2015). In short, the evidence for the state of the early martian climate is mixed, which is consistent with the hypothesis outlined below.

Even if early Mars was generally cold, there are several proposed mechanisms by which the climate could have been warmed transiently, perhaps long enough to form the valleys. (We refer to these as 'cold early Mars' hypotheses). One theory suggests that impacts during the Late Heavy Bombardment created thick steam atmospheres that then rained out to form the valleys (Segura et al., 2012). But the thousands of years of warm climates and the ~600 m of planet-wide rainfall that impacts would have produced are probably several orders of magnitude too short or too low to carve the valleys (Hoke et al., 2011). Extending these warm periods with cirrus clouds (Urata and Toon, 2013) requires 100 percent cloud cover within each cloudy grid cell, combined with a low conversion efficiency of cloud particles to precipitation (their auto-conversion parameter, *B*), both of which seem unlikely. Other GCM studies of $CO_2$-rich early martian atmospheres do not produce warm climates unless additional (unspecified) radiative absorbers are added (Wordsworth et al., 2013, 2015). Alternatively, sporadic volcanic outgassing of $SO_2$ has been suggested as a warming mechanism (Halevy and Head III, 2014), but these authors ignored rainout, as noted earlier.

All of these theories of martian valley formation have overlooked a phenomenon that has been suggested to be important for early Earth, as well as for planets orbiting near the outer edge of their star's habitable zone. Planets on which the $CO_2$ outgassing rate is small (Tajika, 2003), or for which stellar insolation is low (Menou, 2015), are predicted to undergo repeated cycles of global glaciation/deglaciation as a consequence of the dependence of the $CO_2$ removal rate on temperature and $CO_2$ partial pressure, $p$$CO_2$. These 'limit cycles' occur because when the planet is in a glaciated state, $CO_2$ consumption by silicate weathering cannot keep pace with $CO_2$ outgassing from volcanoes. Atmospheric $CO_2$ builds up and increases the planet's surface temperature until it is able to deglaciate. But once the planet is ice-free, $CO_2$ outgassing cannot keep pace with consumption by weathering, so the planet falls back into global glaciation, and the cycle repeats. Such limit cycling does not occur on modern Earth because the solar flux is sufficiently high that weathering can balance outgassing at relatively low atmospheric $p$$CO_2$ (Fig. 1). Furthermore, on an inhabited planet like Earth, soil $p$$CO_2$ is decoupled from atmospheric $p$$CO_2$ by the activities of vascular plants (Berner, 1992). Here we show that on a poorly illuminated early Mars, the high $p$$CO_2$ values required for climatic warmth could have induced rapid weathering, perhaps triggering limit cycles.

## 2. Why climate limit cycles should occur on early Mars but not on Earth

To begin, we looked at terrestrial and martian climates on a global scale using a 1-D climate model coupled to a simple model of silicate weathering. Some important differences between our weathering model and that of Menou (2015) are discussed below. In Section 3, we perform a more complex set of coupled climate/weathering rate calculations using an energy-balance climate model similar to that used by Menou, but with significant updates.

2.1. 1-D climate model calculations for present Earth

Our 1-D climate model has been extensively described elsewhere (Kopparapu et al., 2013; Ramirez et al., 2014), so only a brief summary will be given here. We use the two-stream approximation to treat radiative transfer and assume that the net emitted infrared flux is equal to the net absorbed solar flux in each layer in the stratosphere. Convection in the troposphere is parameterized by assuming a moist $CO_2$ or $H_2O$

**Fig. 1**. **Diagram showing where climate cycles should and should not occur.** The green and red curves represent surface temperatures calculated using a 1D climate model (Kopparapu et al., 2013) for present Earth (green) and early Mars (red) with three different atmospheric compositions and two different surface albedos (0.216 for dry land and ocean, 0.45 for a fully glaciated planet). The solar flux for early Mars is 0.3225 times the flux for present Earth. The brown curve shows the temperature at which the weathering rate balances the present terrestrial $CO_2$ outgassing rate, assuming a $pCO_2^{0.5}$ dependence (solutions for Eqn. 3 when $W/W_{\oplus} = 1$). The blue circle shows the location of the present soil pCO2 level. These curves define three regions of climate stability: 1) *Warm stability*: The surface temperature and weathering rate curves intersect above the freezing point (abiotic Earth) and planets remain permanently de-glaciated 2) *Cold Stability*: The surface temperature fails to ever rise above the freezing point of water (red dotted early Mars case). Such planets remain permanently glaciated. 3) *Limit Cycling*: The surface temperature and weathering rate curves intersect below the freezing point, but temperatures above freezing are possible as $CO_2$ and $H_2$ build up (solid/dashed red early Mars case with $H_2$).

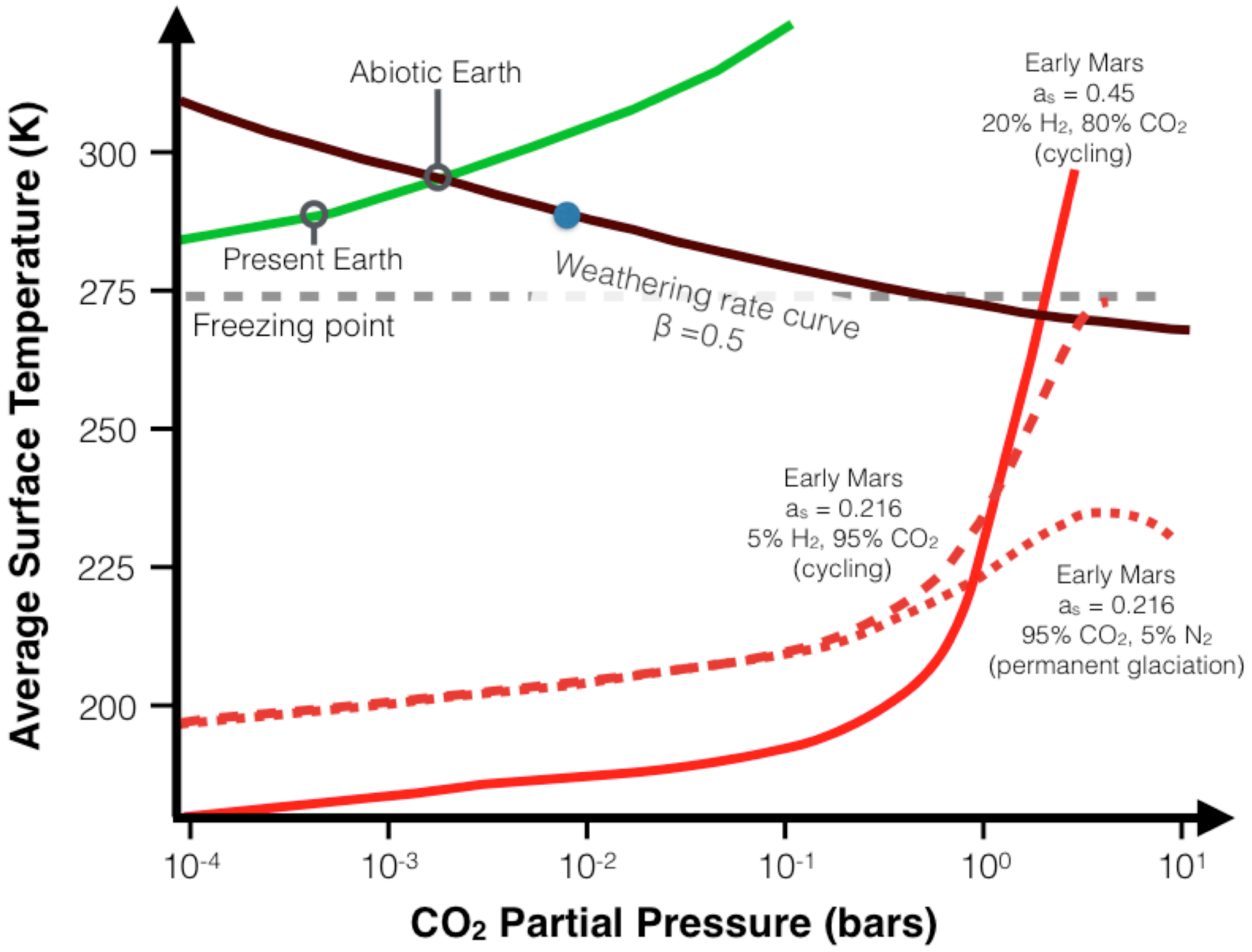

adiabatic lapse rate, following (Kasting, 1991).

Clouds are not included explicitly in our 1D model, but their effect on climate is simulated by adjusting the surface albedo, $a_s$. For Earth, the actual surface albedo is ~0.1 or below, but we use a value of 0.31. This allows the model to reproduce the present mean surface temperature of 288 K, given present solar insolation and present (or preindustrial) $pCO_2$, $3\times10^{-4}$ bar. This calculation is shown by the point labeled 'Present Earth' in Fig. 1. As $pCO_2$ is increased, the surface temperature increases along the green curve. Our model predicts about 10 K of warming per factor of 10 increase in $CO_2$, or 3 K per $CO_2$ doubling, which is near the middle of the 1.5-4.5 K range predicted using more complicated 3-D climate models (IPCC, 2013).

We have done 1-D climate model calculations for early Mars, as well. Before discussing those, though, we describe the other part of our coupled model, namely, the dependence of $p\mathrm{CO}_2$ on surface temperature via weathering.

2.2 The Carbonate-Silicate Cycle:

$p\mathrm{CO}_2$ is a function of surface temperature by way of its participation in the carbonate-silicate cycle (Menou, 2015; Tajika, 2003). We modeled the time-dependent mass exchange of $CO_2$ with the crust in a manner similar to that of Menou:

$$\frac{d}{dt}(pCO_2) = V - W \quad [1]$$

Here, $p\mathrm{CO}_2$ is the partial pressure of $CO_2$ in bar, $V$ is the volcanic outgassing of $CO_2$ (in bar/Gyr), and $W$ is the rate at which $CO_2$ gets removed via rock weathering. These outgassing units are non-standard and can lead to confusion. Menou took the $CO_2$ outgassing rate for Earth to be 7 bar/Gyr. We can convert this rate to geochemists' units of Tmol/yr ($10^{12}$ mol/yr) using the relationship:

$$P_s = M_{col} \cdot g \quad [2]$$

Here, $P_S$ is surface pressure, $M_{col}$ = column mass, and $g$ = gravitational acceleration. Calculating the column mass for a 1-bar atmosphere, multiplying by the surface area of the Earth, and converting to moles yields the relationship: 1 bar $CO_2$ $=10^5$ Pa $CO_2$ = $1.18\times10^{20}$ moles. Using this conversion factor, Menou's outgassing rate comes out to be 0.83 Tmol/yr. This is about a factor of 10 lower than the estimates for the terrestrial $CO_2$ outgassing rate, ~7.5 Tmol/yr (Gerlach, 2011; Jarrard, 2003 and refs therein)

The relationship between $p\mathrm{CO}_2$ and $CO_2$ atmospheric mass is different on Mars because gravity and surface area are both lower by a factor of 3.6 (for gravity) to 3.5 (for surface area). These factors offset each other, so the result is: 1 bar $CO_2$ is ≅ $1.18\times10^{20}$ moles ·(0.75) = $8.85\times10^{19}$ mol. Ramirez et al. (2014) have suggested that outgassing rates per unit area on early Mars might be similar to those on present Earth based on the fact that heat flow per unit area is thought to be similar on the two planets (Zuber et al., 2000). Halevy and Head (2014) make similar assumptions. If so, the global outgassing rate on early Mars would be scaled by surface area relative to Earth, yielding 7.5 Tmol/yr · (0.28) ≅ 2.1 Tmol/yr ≅ 24 bar/Gyr. In the calculations described in Sect. 2.3, we have let the $CO_2$ outgassing rate be a free parameter, and we have varied its value from 0.8-4.5 Tmol/yr, spanning the region of parameter space near the modern terrestrial value. As pointed out earlier, in order for outgassing rates on early Mars to be comparable to those on modern Earth, the initial volatile inventory must have been large, and volatiles must have been recycled between the crust and the mantle.

According to eq. (1), the silicate weathering rate, $W$, must balance the volcanic outgassing rate, $V$, when the system is in steady state. Following Berner (1994), we write the weathering rate as:

$$\frac{W}{W_\oplus} = \left(\frac{pCO_2}{p_\oplus}\right)^\beta e^{k_{act}(T_{surf}-288)}\left[1 + k_{run}\left(T_{surf} - 288\right)\right]^{0.65}. \quad [3]$$

This equation was formulated for the (biotic) modern Earth and, so, must be scaled to a presumably abiotic early Mars. The parameter $k_{act}$ = 0.09 $K^{-1}$ is an activation energy, and $k_{run}$ = 0.045 $K^{-1}$ is a runoff efficiency factor. $W_\oplus$ and $p_\oplus$ are the weathering rate and effective $CO_2$ partial pressure on modern Earth (see discussion below).

Extending our analogy between the outgassing on early Mars and the outgassing on present Earth, we assume that $W_\oplus \equiv V_\oplus$ = 2.1 Tmol/year for $T_{surf}$ = 288 K.

The parameter, β, dictates the dependence of the weathering rate on $pCO_2$. The value of β is uncertain, even when weathering is not influenced by biology. Values of β between 0 and 1 are theoretically possible, depending on whether one has open or closed system weathering (Berner, 1992). Walker et al. (1981) assumed β = 0.3 based on laboratory experiments by Lagache (1976). If the weathering rate is proportional to the dissolved $[H^+]$ in groundwater, then β should be equal to 0.5. We assumed β = 0.5 in all of our calculations.

The parameter $p_\oplus$ is complicated. Menou used $p_\oplus = 3.3\times10^{-4}$ bar, which is approximately the preindustrial atmospheric value. But weathering occurs in soils, and soil $pCO_2$ on Earth is 10-100 times higher than atmospheric $pCO_2$ as a consequence of root respiration by vascular plants (Kump et al., 2010). Here, we take the geometric mean of these two values and assume that soil $pCO_2$ is 30 times atmospheric $pCO_2$. When combined with an assumed β value of 0.5, this implies that land plants accelerate silicate weathering by a factor of $30^{0.5} \cong 5.5$. By comparison, Berner, (1994) assumed that weathering on the pre-land-plant Earth was slower by a factor of 0.15, implying that land plants enhance weathering by a factor of (1/0.15) ≅ 6.7. Given the large uncertainties in this process, the agreement between these estimates is pretty good. We should note that biological influences on weathering are not limited to enhancing soil $pCO_2$. Plants also produce organic acids that are more effective than carbonic acid at dissolving rocks. Consequently, some authors (e.g., Schwartzman and Volk, 1989) have argued that land plants accelerate silicate weathering by as much as a factor of 1000. We follow Berner and adopt a more conservative estimate, but we acknowledge that significant uncertainty remains as to exactly how much weathering is affected by the biota.

Following this reasoning, we set $p_\oplus$=0.01 bar in eq. (3), consistent with the current soil $pCO_2$ value. This implies that $W/W_\oplus = 1$ at the point ($pCO_2$, $T_{surf}$) = (0.01 bar, 288 K). This point is shown as a blue circle along the weathering rate curve in Fig. 1.

Also shown at the intersection of the weathering rate curve and the (green) greenhouse-warming curve is a point labeled 'Abiotic Earth'. This represents the ($pCO_2$, $T_{surf}$) values that the Earth would be expected to achieve if life were suddenly eliminated. It is calculated as follows: With β = 0.5, reducing $p_\oplus$ from 0.01 bar to $3.3\times10^{-4}$ bar reduces the weathering rate by a factor of $30^{0.5} \cong 5.5$. That's how much the weathering rate would decrease if life were instantaneously eliminated. Both $T_{surf}$ and $pCO_2$ would then gradually increase until the weathering rate went back up to its original value. This happens at ($pCO_2$, $T_{surf}$) ≅ (0.002 bar, 295 K). Thus, our model predicts that present Earth would be about 7 degrees warmer if land plants were to suddenly disappear.

Whether or not limit cycling should occur depends on where the surface temperature curve (green curve for Earth) intersects the weathering rate curve. If the intersection is *above* the freezing point of water, as it is for modern Earth (or for the abiotic Earth), then stable, warm solutions are possible. If the intersection point occurs *below* the freezing point, then either the climate system is subject to limit cycles or the surface remains permanently frozen. Note that the actual weathering rate cannot follow the weathering rate curve in this region, at least in a globally averaged model, because weathering should essentially cease at surface temperatures below the freezing point of water. This makes it even more obvious that steady-state solutions are not possible in this region of parameter space.

### 2.1. 1-D climate model calculations for early Mars

This brings us, finally, to Mars. For Mars, a surface albedo, $a_s$, of 0.216 allows the model to reproduce the present mean surface temperature of 218 K, given the present martian solar flux (0.43 times that of Earth), along with a fully saturated, 95% $CO_2$, 5% $N_2$, 6-mbar atmosphere (calculations

not shown). The model was then run with this same value of $a_s$ for early martian conditions (3.8 Gyr ago) when the solar luminosity was 75 percent of present (Fig.1, red dotted curve). Not surprisingly, the calculated mean surface temperature never reaches the freezing point of water, regardless of how much $CO_2$ is added. Instead, it turns over and begins decreasing at $pCO_2 > \sim3$ bar. This result is expected, as virtually all climate models predict that early Mars could not have been warmed to the freezing point by a $CO_2$-$H_2O$ atmosphere.

We can produce warmer surface temperatures by adding $H_2$ to the early martian atmosphere, following Ramirez et al. (2014). Ramirez et al. showed that above-freezing surface temperatures for early Mars are attainable with as little as 3 bar of $CO_2$ and a 5% mixing ratio of $H_2$ (red dashed curve, Fig.1). A 3-bar atmosphere is within the initial inventory estimated earlier for a volatile-rich early Mars. As can be seen from the figure, however, this curve intersects the weathering rate curve *below* the freezing point of water. As pointed out above, this implies that no stable steady state is possible. Instead, the climate should oscillate between cold, globally glaciated conditions and warmer, ice-free conditions.

$H_2$ mixing ratios exceeding 5% are needed to produce warm climates in our EBM simulations, described below, because the planet needs to be able to deglaciate, starting from a state with a frozen surface and high albedo. For the red solid curve in Fig. 1, $a_s = 0.45$ which is consistent with a frozen surface (see Section 3.1.1). In this case, ~20% $H_2$ is needed to deglaciate the planet. The intersection of this red curve with the weathering rate curve is again below the freezing point of water, indicating that limit cycles should still occur. We explore these limit cycles quantitatively in the next section, using a more complicated climate model.

## 3. Exploring climate limit cycles on early Mars with an EBM

1-D climate model calculations are useful for determining when climate limit cycles should occur. To explore these cycles, though, it is appropriate to use an energy-balance climate model (EBM) similar to those used by Menou (2015) and Haqq-Misra et al. (2016) because EBMs can simulate ice-albedo feedback. Menou's model was actually a zero-dimensional parameterization of an EBM. Below we describe our own, 1-D (in latitude) EBM, also used in Haqq-Misra et al. (2016), and how we used it to look at early martian climate cycles.

### 3.1 Energy-Balance Climate Model Description:

Our EBM, like the one parameterized by Menou, was originally created by Williams and Kasting (Williams, 1997). We updated this EBM with new radiative transfer parameterizations from our 1-D model. Details concerning the EBM are included in the following subsections and the parameterizations are included in Supp. Info. Table 1 shows that numerous parameters are included within an EBM.

In addition to calculating surface temperature as a function of latitude, our EBM simulates geochemical cycles of the greenhouse gases, $CO_2$ and $H_2$. The $CO_2$ cycle was described in the previous section. The only difference is that silicate weathering occurs at a range of different temperatures corresponding to different latitude bands. These individual rates are combined in an area-weighted manner to yield the global weathering rate. The atmospheric $H_2$ concentration is determined by balancing volcanic outgassing with diffusion-limited escape to space (see Section 3.1.2). Escape slows down as the atmosphere becomes denser; thus, the concentrations of $H_2$ and $CO_2$ are positively correlated. The outgassing rates of $H_2$ and $CO_2$ are treated as free parameters in the modeling exercise.

#### *3.1.1. Basic model equations*

Our EBM calculates the meridionally averaged temperature as a function of latitude and time according to

$$C\frac{\partial T}{\partial t} = \bar{S}(1-\alpha) - F_{OLR} + \frac{1}{\cos\theta}\frac{\partial}{\partial\theta}\left(D\cos\theta\frac{\partial T}{\partial\theta}\right) \quad [4]$$

Here, $\theta$ is latitude and $t$ is time. $\bar{S} = S \cdot q(\theta)$ is the diurnally averaged solar flux, which accounts for seasonal changes in the solar flux ($S$) at each latitude band due to seasonal changes. $C$ is the effective heat capacity of the surface. It is calculated by the procedure described in (Fairén et al., 2012; Williams, 1997). $D$, the diffusion coefficient, accounts for the energy transport across latitude bands. It is scaled for a present-day Earth (Fairén et al., 2012) and depends on changes in surface pressure, atmospheric mass and heat capacity, and rotation rate.

The original radiative transfer parameterizations, from Williams and Kasting (1997), were replaced for this study with new parameterizations based on the 1-D model of Kopparapu et al. (2013). Over 40,000 calculations with the 1-D model were used to investigate the dependence of radiative fluxes on different parameters. The outgoing long wave flux, $F_{OLR}$, was calculated as a $4^{th}$-order polynomial in surface pressure, surface temperature, and volume mixing ratio of $H_2$. The top-of-atmosphere albedo, α, (which is related to the absorbed solar flux) was split into two $3^{rd}$-order polynomials. One is for surface temperatures less than 250 K and the other is for surface temperatures between 250 K and 350 K. Both fits are calculated as a function of zenith angle, surface pressure, surface temperature, $H_2$ volume mixing ratio and surface albedo. Both the OLR and albedo parameterizations assume early Mars' solar constant and Mars' radius. They also assume an $H_2O$-$H_2$-$CO_2$ atmosphere.

Surface albedo is calculated at each latitude band as a weighted sum of unfrozen land, unfrozen ocean, and fractional ice coverage. Unfrozen land is given a fixed albedo of 0.216, while the ocean albedo is allowed to vary, depending on solar zenith angle. $H_2O$ ice is given a fixed albedo of 0.45, consistent with slightly dirty ice. The $CO_2$ ice albedo is given a fixed value of 0.35, but is only used when $CO_2$ is condensing onto the surface, which does not occur in the following calculations. (Warren et al., 1990). If the partial pressure of $CO_2$ exceeds the saturation vapor pressure at the corresponding surface temperature, we assume it condenses on the surface as dry ice over the water ice. We also do not let the surface temperature fall below the saturation temperature of $CO_2$. If it does fall below, we bring the temperature up to the saturation temperature. Similar behavior is observed during winter above the martian polar caps today.

*3.1.2 Hydrogen escape*

As mentioned previously, molecular hydrogen ($H_2$) is required to bring the early martian surface temperature above freezing. We assume that escape of $H_2$ is diffusion limited (Walker, 1977) and obeys the equation:

$$\phi_{esc}(H_2) = \frac{b_i}{H_a}\frac{f_{H_2}}{1+f_{H_2}}. \quad [4]$$

Here, $\frac{b_i}{H_a} = 5\times10^{11}$ $cm^{-2}s^{-1}$ for early Mars (Batalha et al., 2015; Pater and Lissauer, 2001). For each time step, the escape rate of $H_2$ is calculated. We then increment the column density of $H_2$ according to:

$$\frac{dn_{col}(H_2)}{d\tau} = \phi_{out}(H_2) - \phi_{esc}(H_2). \quad [5]$$

Here, $\phi_{out}(H_2)$ is the volcanic outgassing rate of $H_2$ per unit area. The column density of $CO_2$ is related to its time varying partial pressure via

$$n_{col}(CO_2) = pCO_2 / (g \cdot m_{CO_2}) \quad [6]$$

Here, $g$ is gravitational acceleration and $m_{CO2}$ is the molecular mass of carbon dioxide. After each time step, the mixing ratio of hydrogen is recalculated from the column density of $H_2$ and $CO_2$:

$$f_{H_2} = n_{H_2} / (n_{col}(CO_2) + n_{H_2}). \quad [7]$$

Because $n_{CO2}$ is proportional to $pCO_2$, a decrease in the $CO_2$ partial pressure will lead to an increase in the $H_2$ mixing ratio, and hence to an increase in the $H_2$ escape rate, following eq. (4).

*3.1.3 EBM modeling results*

To explore the effect of limit cycles on early martian climate, we performed over 30

calculations with the EBM, using $CO_2$ outgassing rate and $H_2$ outgassing rate as free parameters. The results are summarized in Fig. 2. The bottom panel shows the fraction of time within a billion year period when at least one latitude band was above freezing (273 K). When the $H_2$ outgassing rate is <~400 Tmol/year, cycling does not occur because there is not enough greenhouse warming to deglaciate the frozen planet. This rate is over 150 times higher than the $H_2$ outgassing rate on modern Earth, but is not out of line for a tectonically active early Mars with a highly reduced mantle (Batalha et al., 2015; Chassefière et al., 2014). This rate could be reduced if hydrogen escaped at less than the diffusion limit. The $CO_2$ outgassing rates studied ranged from 0.8 Tmol/yr up to about 4.5 Tmol/yr. (The equivalent modern Earth rate, when adjusted for the difference in planet surface area, is 2.1 Tmol/yr). Low $CO_2$ outgassing rates create low-frequency cycles. High $H_2$ outgassing rates decrease the length of the warm phase of each individual cycle. Panel 'A' shows typical cycling behavior for moderate $H_2$ outgassing rates and low $CO_2$ outgassing rates. Limit cycles provide warm periods lasting up to 5-10 Myr, separated by long (160 Myr) periods of below-freezing temperatures. Panel 'B' shows what happens when outgassing rates are too low to induce cycling. Panel 'C' shows a case in which cycling is occurring at such high frequency that the planet is warm nearly 50% of the time.

**Figure 2.** The heat map shows the fraction of time within a billion year period when the maximum surface temperature was above freezing (273 K). Panel 'A', 'B', and 'C' show cases where limit cycles are moderate, absent, and rapid, respectively. For reference, the modern $CO_2$ and $H_2$ outgassing rates on present Earth are 7.5 Tmol/year (Jarrard, 2003) and 2.4 Tmol/year (Ramirez et al., 2014 and refs. therein), respectively. These rates are scaled down by a factor of 3.5 for Mars to account for its smaller surface area.

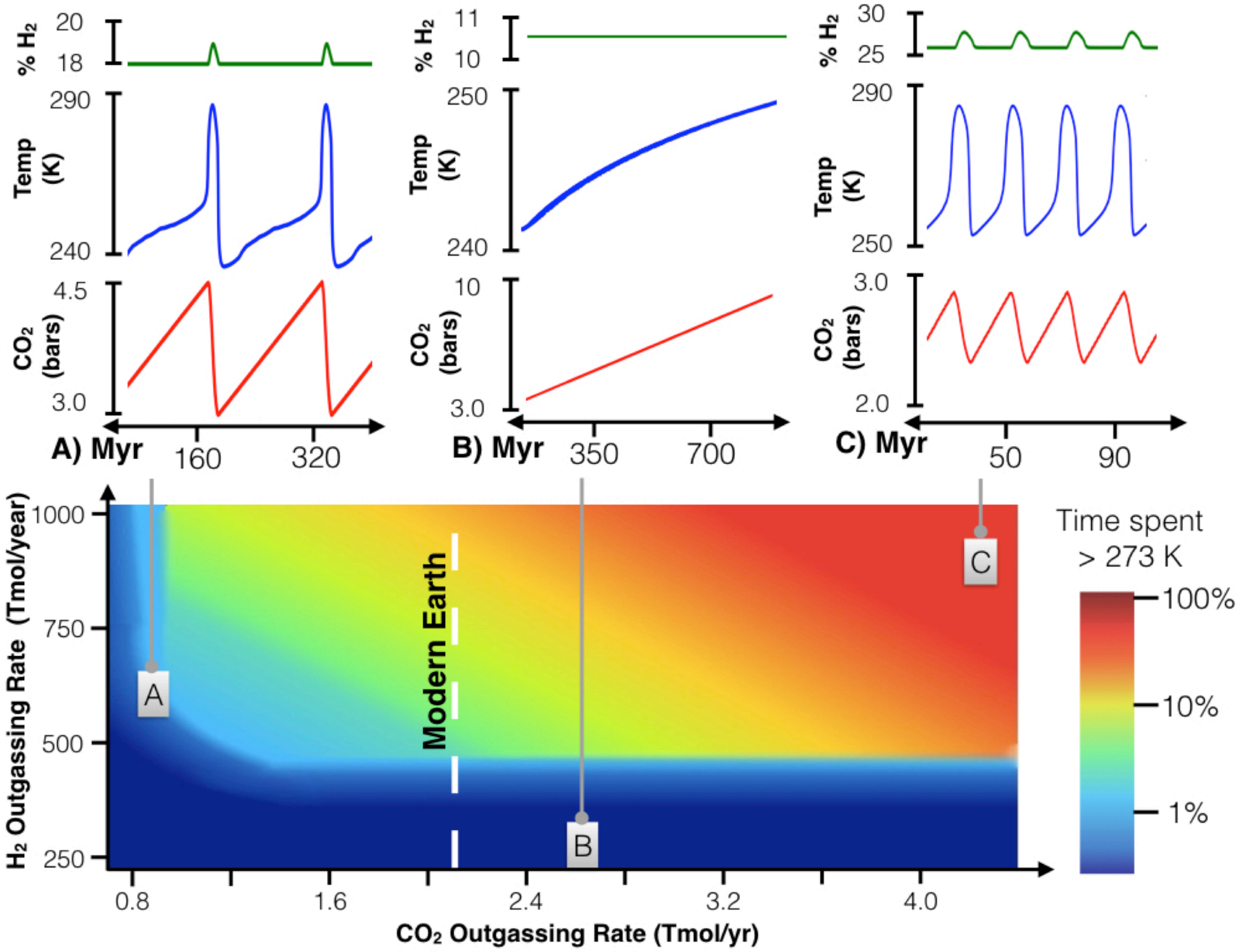

Our favored region of parameter space to approximate the actual climate of early Mars is in the area of panel 'A'. 10 Myr is an adequate time period to form the larger martian valleys, according to the references mentioned in Sect. 1. And the fact that the surface remains frozen for much longer time periods may help to reconcile our results with those who have favored the 'cold early Mars' hypothesis.

## 4. Discussion

Some of the parameters included in Table 1 are not well defined for early Mars, while others have well documented values. An ocean fraction of 20% is consistent with the recent five-year monitoring of atmospheric D/H ratios (Villanueva et al., 2015). A northern ocean land configuration is consistent with observations of deltas and valleys (Di Achille and Hynek, 2010), as well as observations of the martian topography (Perron et al., 2007). A present day martian obliquity of 25° was chosen for most runs but we also explored cases where obliquity was chaotically varying from 25-50° (see discussion below) (Touma and Wisdom, 1993).

### 4.1. Sensitivity to outgassing rates and ice albedo

The parameters that are less well documented are $CO_2$ and $H_2$ outgassing rates and $H_2O$ ice albedo. Getting constraints on outgassing rates is challenging. Estimates made by looking at the igneous rocks and making assumptions about the volatile content of the initial lava from which they formed are universally too low to produce a warm early martian climate (Batalha et al., 2015). Some type of volatile recycling mechanism, such as plate tectonics, is required to do this. As discussed in the main text, high $H_2$ outgassing rates are expected because of Mars' highly reduced mantle and the presence of ultramafic rock on its surface (Chassefière et al., 2014).

Ice albedo is a critical parameter in our model. Fig. 2 shows one case where 18% $H_2$ is needed to sustain climate cycles. This value can be lowered by decreasing the water ice albedo. Dirtier ice would increase the amount of absorbed solar radiation. High ice albedos encourage limit cycling but require high concentrations of $CO_2$ and $H_2$ to deglaciate. Very low ice albedos (< 0.36) disrupt the limit cycling behavior. In most of our simulations, we keep the water ice albedo fixed at a relatively high value of 0.45 and vary $H_2$ and $CO_2$ outgassing rates. Doing so produces the heat map shown in Fig. 2.

### 4.2. Sensitivity to obliquity

The martian climate is also sensitive to obliquity, which is thought to have varied chaotically throughout Mars' lifetime (Laskar et al., 2004). We looked at this by changing the assumed obliquity from 25° to 50° in the EBM. The effects of obliquity interact with the assumed continental distribution. High obliquity causes more incident solar radiation at the poles and less at the equator. With our specification of a northern ocean, high obliquity increases the frequency of cycles, but not the length of each individual cycle. This admittedly does not exhaust the list of possible combinations of obliquity and geography. However, on the basis of this experiment, we infer that higher obliquity would have encouraged limit cycling rather than inhibiting it.

We also did a second simulation in which the obliquity was allowed to cycle repeatedly between $25^0$ and $50^0$ for 1 Gyr. The limit cycles were found to occur at the frequency corresponding to the higher obliquity state. Our specified value of $25^0$ is thus conservative; higher values would lead to even more pronounced limit cycling behavior.

## 5. Conclusion

The duration of the warm periods caused by limit cycles is comparable to that needed to form the larger martian valleys, $10^6$-$10^7$ yrs (Hoke et al., 2011). The length and the frequency of cycling behavior depend on the assumed $CO_2/H_2$ outgassing rates and on the planet's obliquity. For higher outgassing rates, the length of each warm phase decreases but their frequency increases. Smaller outgassing rates give longer, more widely spaced warm periods, which may be more consistent with the timescales needed for valley formation. If this type of climate cycling was indeed occurring on early Mars, then both the warm and the cold early Mars hypotheses are partly correct. MSL and future Mars missions may be able to test this hypothesis by refining the time scales for valley and lake formation and by determining whether multiple warming events took place.

**Acknowledgments:** The authors thank Darren Williams for assistance with model development. This material is based upon work supported by the National Science Foundation under Grant No. DGE1255832 to N.E.B. J.H. acknowledges funding from the NASA Habitable Worlds program under award NNX15AQ82G. R.K.K. and J.F.K acknowledge funding from NASA Astrobiology Institute's Virtual Planetary

Laboratory lead team, supported by NASA under cooperative agreement NNH05ZDA001C. Any opinions, findings, and conclusions or recommendations expressed in this material are those of the author(s) and do not necessarily reflect the views of NASA or the National Science Foundation.

**Table 1. Parameters for input into EBM.**

| Model Parameters | Value or range of |
|---|---|
| **Eccentricity** | 0 |
| **Obliquity** | 25-50 |
| **Surface Pressure** | 3 bars |
| **Ocean Coverage** | 20% |
| **Geography** | Northern ocean |
| **Solar Constant relative** | 0.75 |
| **$CO_2$ outgassing rate** | 2.1 Tmol/year |
| **$H_2$ outgassing rate** | $1\times10^{12}$ - $4.5\times10^{12}$ $cm^{-2}s^{-1}$ |
| **Ground Albedo** | 0.216 |
| **$H_2O$ Ice Albedo** | 0.38-0.66 |
| **$CO_2$ Ice Albedo** | 0.35 |
| **β (Eqn 1)** | 0.5 |
| **$K_{act}$ (Eqn 1)** | 0.09 |